\documentclass{aa}
\usepackage{graphicx}

\def\simge{\mathrel{    \rlap{\raise 0.511ex \hbox{$>$}}{\lower
0.511ex \hbox{$\sim$}}}}
\def\simle{\mathrel{    \rlap{\raise 0.511ex \hbox{$<$}}{\lower
0.511ex \hbox{$\sim$}}}}

\begin{document}

\title{The high energy gamma-ray emission expected from Tycho's supernova remnant}

   \subtitle{}

   \author{H.J.V\"olk
          \inst{1}
          \and         
            E.G. Berezhko
          \inst{2}
           \and
          L.T.Ksenofontov
          \inst{2}
           \and
           G.P. Rowell
          \inst{1}}

   \offprints{H.J.V\"olk}

   \institute{Max Planck Institut f\"ur Kernphysik,
                Postfach 103980, D-69029 Heidelberg, Germany\\
              \email{Heinrich.Voelk@mpi-hd.mpg.de}
             \email{Gavin.Rowell@mpi-hd.mpg.de}
         \and
                 Institute of Cosmophysical Research and Aeronomy,
                     31 Lenin Ave., 677891 Yakutsk, Russia\\
              \email{berezhko@ikfia.ysn.ru}
              \email{ksenofon@ikfia.ysn.ru}
             }

   \date{Received month day, year; accepted month day, year}

     \abstract
{A nonlinear kinetic model of cosmic ray (CR) acceleration in
supernova remnants (SNRs) is used to describe the properties of Tycho's SNR.
Observations of the expansion characteristics and of the nonthermal radio and
X-ray emission spectra, assumed to be of synchrotron origin, are used to
constrain the overall dynamical evolution and the particle acceleration
parameters of the system, in addition to what is known from independent
estimates of the distance and thermal X-ray observations.  It is shown that a
very efficient production of nuclear cosmic rays, leading to strong shock
modification, and a large downstream magnetic field strength
$B_\mathrm{d}\approx240$~$\mu$G are required to reproduce the observed
synchrotron emission from radio to X-ray frequencies. This field strength is
still well within the upper bound for the effective magnetic field, consistent
with the acceleration process. The $\pi^0$-decay $\gamma$-ray flux turns out to
be somewhat greater than the inverse Compton (IC) flux off the Cosmic Microwave
Background at energies below 1 TeV, dominating it strongly at 10 TeV.  The
predicted TeV $\gamma$-ray flux is consistent with but close to the very low
upper limit recently obtained by HEGRA.  A future detection at
$\epsilon_{\gamma} \sim 10$~TeV would clearly indicate hadronic emission.

    }
   \maketitle
%

\section{Introduction}

The observations of Tycho's supernova remnant (SNR) with the HEGRA
stereoscopic system of imaging atmospheric Cherenkov telescopes (IACTs) on
La Palma have recently been analyzed. The object has long been considered
as a prototype candidate hadronic CR source in the Northern Hemisphere
(e.g.  Drury et al. \cite{dav94}), although it was always clear that the
sensitivity of the present generation of IACTs is marginal for a
detection. In fact, after $\sim 65$ hours of observation time, HEGRA did
not detect Tycho's SNR, but it could establish a very low $ 3 \sigma$
upper flux limit of $5.78 \times 10^{-13}$ photons cm$^{-2}$ s$^{-1}$, or
33 milli-Crab, at energies $ > 1$~TeV (Aharonian et al. \cite{aha01}).
This value is about a factor 4 lower than the one previously published by
the Whipple collaboration (Buckley et al. \cite{buc98}), assuming a
spectral index of $ - 1.1$ for the comparison. In the above HEGRA paper on
Tycho the existing radio and X-ray synchrotron observations were used to
infer a lower limit to the mean magnetic field strength in the remnant due
to the nondetection of Inverse Compton (IC) emission off the Cosmic
Microwave Background (CMB).  At the same time, published estimates of the
hadronic $\pi^0$-decay gamma-ray emission (Berezhko \& V\"olk \cite{bv97})
were employed and scaled to the parameters of Tycho's
SNR, to compare with the new upper flux limit. The predictions from the
time-dependent kinetic model (Berezhko et al. \cite{byk96})  were also
renormalized to take the expected deviations from spherical symmetry of
the nucleon injection rate for a Type Ia SNR into account, and were found
to be consistent with the present HEGRA nondetection in $\gamma$-rays,
although the predicted flux values for the hadronic emission were only
slightly smaller than the observational upper limit.

This tantalizing situation has prompted us to model the acceleration of
both electrons and protons in detail with the nonlinear kinetic theory,
using the observed synchrotron emission as a constraint on the electron
acceleration characteristics, and thereby to model the hadronic and IC
$\gamma$-ray emission simultaneously.

We demonstrate that, together with the renormalization, the existing data
are consistent with very efficient acceleration of CR nuclei at the SN
shock wave which converts a significant fraction of the initial SNR energy
content into CR energy.  This energy is distributed between energetic
protons and electrons in a proportion similar to that of the Galactic CRs.
Therefore Tycho's SNR might indeed be a typical Galactic CR source. Direct
evidence for the inferred strong production of nucleonic CRs in Tycho
would not only require a $\gamma$-ray detection as such, but in particular
a spectrum measurement, ideally beginning at photon energies of about 100
MeV up to the highest energies. Of particular importance would be
measurements in excess of 10 TeV particle energy which are beyond the
reach of the accelerated electron component that suffers a cutoff at
energy of about 50~TeV due to synchrotron losses in our model.  This {\it
specific spectral behavior} implies a high magnetic field strength of
about 40~$\mu$G upstream of the outer SNR shock. Such a field strength is
indeed required to fit the nonthermal radio and X-ray synchrotron
emission. We shall give an upper bound on the effective magnetic field in the 
discussion section. It well encompasses the deduced field strength.

\section{Model}

A supernova (SN) explosion ejects a shell of matter with total energy
$E_\mathrm{sn}$ and mass $M_\mathrm{ej}$.  During an initial period the
shell material has a broad distribution in velocity $v$. The fastest part
of these ejecta is described by a power law $dM_\mathrm{ej}/dv\propto
v^\mathrm{2-k}$ (e.g. Jones et al.  \cite{Jones81}, Chevalier
\cite{chev82}). The interaction of the ejecta with the interstellar medium
(ISM) there creates a strong shock which accelerates particles.

Our nonlinear model (Berezhko et al. \cite{byk96}; Berezhko \& V\"olk
\cite{bv97})  is based on a fully time-dependent solution of the CR
transport equations together with the gas dynamic equations in spherical
symmetry. Regarding the coefficients in the transport equations and the
boundary conditions as well as the approximation of spherical symmetry, we
shall introduce a number of approximations as follows (see also the
discussion in Berezhko et al. \cite{bkv02}):

The CR diffusion coefficient is taken as the Bohm limit 
\begin{equation}
\kappa (p)=\kappa (mc) (p/mc), 
\end{equation}
where $\kappa(mc)=mc^2/(3eB)$, $e$ and 
$m$ are the particle charge and
mass, $p$ denotes the particle momentum, $B$ is the magnetic field
strength, and $c$ is the speed of light. This limiting value is consistent
with the high, probably turbulently amplified magnetic field strength
(Lucek \& Bell \cite{lucb00}) due to the strong pressure gradient of the
energetic particles in the very strong shock of the young remnant. In
fact, Eq. (1) uses the relativistic form of Bohm diffusion, equal for both
electrons and nuclei. In fact, the actual form of $\kappa (p)$ at low
momenta $p \ll p_\mathrm{max}$ does not influence the particle spectrum if
$\kappa (p)$ is a strongly increasing function of $p$. In addition at
nonrelativistic proton energies $p < m_\mathrm{p} c$, even if the electron
and proton diffusion coefficients are different at the same $p$, their
spectra $f(p)$ are essentially the same because these low energy particles
do not produce any shock modification and therefore their spectrum is
$f(p) \propto p^{-q_\mathrm{s}}$, where $q_\mathrm{s}$ is the spectral
index appropriate for the subshock compression ratio $\sigma_\mathrm{s}$,
calculated below.

The number of suprathermal protons injected into the acceleration process
is described by a dimensionless injection parameter $\eta$ which is a
fixed fraction of the ISM particles entering the shock front. For
simplicity it is assumed that the injected particles have a velocity four
times higher than the postshock sound speed. It is expected that ion
injection is quite efficient at the quasi-parallel portion of the shock
surface, where it is characterized by the value $\eta \sim 10^{-4}$ (see
V\"olk et al. \cite{voelk} for details). In momentum space the accelerated
(nonthermal) proton distribution function grows smoothly out of the
downstream thermal distribution (e.g. Malkov \& V\"olk \cite{malkv95}).

Assuming nucleon injection to occur through the leakage of suprathermal
particles into the upstream region from behind the shock, the relevant
injection velocity parallel to the magnetic field is $V_{\mathrm {s}}
\mathrm {cos}^{-1}\phi$, where $V_{\mathrm {s}}$ denotes the (radial)  
shock velocity and $\phi$ is the angle between the shock normal and the
{\it downstream} magnetic field direction; it is refracted away from the
shock normal. Therefore the injection is expected to be strongly
suppressed at the quasi-perpendicular surface fraction of the shock and
this surface fraction is larger than that of the quasi-parallel fraction
due to the field refraction. This lack of symmetry in the actual SNR can
be approximately taken into account by a renormalization factor
$f_\mathrm{re}<1/2$ which diminishes the nucleonic CR production efficiency as
calculated in the spherical model, and all effects associated with it. A
more detailed estimate, taking also the diffusive broadening of the
quasi-parallel area into account, yields roughly $f_\mathrm{re}=0.15$ to 0.25 
(V\"olk et al. \cite{voelk}).

This is not necessarily in conflict with the radio polarization result of
Reynoso et al. (\cite{rmg97}) and Dickel et al. (\cite{di91}) who obtain
from their VLA measurements a roughly radial ordered magnetic field
component over most of the shock surface. The fractional polarization of
10\% to 15\% is rather small (Dickel et al. \cite{di91}), indicating a
rather small excess of ordered radial magnetic field (see also the
discussion in Reynolds \& Gilmore \cite{rg93}). Obviously Tycho's SNR is
not a magnetic monopole. To resolve this puzzle we argue that there exist
at least two physical processes which tend to produce radial field
directions in a remnant that is still in the sweep-up phase and as a whole
embeded in a uniform ISM and magnetic field. First of all, such young SNRs
as Tycho's are Rayleigh-Taylor unstable during the sweep-up phase (Gull
\cite{gul73}; Vel\'azquez et al. \cite{vgd98}).  Secondly, the ejecta
themselves may be so nonuniform that they can wrinkle the blast wave
surface (Aschenbach et al. \cite{ash95}). In addition, the ambient ISM
around Tycho appears to be unexpectedly nonuniform as well, cf. Reynoso et
al. (\cite{rmg97}). Even though these irregularities are of a macroscopic,
hydromagnetic nature, their size spectrum probably extends to quite small
scales. The acceleration scales can even be much smaller.  In fact, the
acceleration scales of protons with multi-GeV energies, on whose strong
self-excited magnetic irregularities (Bell \cite{bell78}) the electrons
may parasitically accelerate as well, is of order $10^{-4}$ times the
shock radius or even smaller, and the proton scales are connected with
quasi-parallel, i.e. quasi-radial field line portions. On other portions
no acceleration occurs, and they will not be illuminated by synchrotron
emitting electrons. Thus it is possible that the magnetic field appears
essentially radial over the remnant surface even to the VLA. We leave it
open here to which extent small-scale magnetic irregularities that lead to
measurable polarization effects can also be nonlinearly produced by the
accelerated particle population itself.

Due to their small mass $m_{\mathrm {e}}$, suprathermal electrons cannot
resonantly interact with the hydromagnetic wave turbulence created by the
energetically dominant accelerating ions. Therefore electrons can only
participate in the diffusive shock acceleration process at high energies
$\leq m_{\mathrm {e}}c^2$ , albeit still nonrelativistic as far as nuclei
are concerned. Presumably they are injected by electrostatic fluctuations
(e.g. Malkov \& Drury \cite{malkd}). For simplicity we can consider their
acceleration to start from the same momentum as protons start, if we
restrict our further consideration to relativistic electrons, at energies
possibly far above their actual injection energy. What counts alone for
the considerations in this paper is that at energies corresponding to
those of relativistic protons - only those energies play a role for the
observed synchrotron emission - electrons have exactly the same dynamics
as the protons. Therefore, below the momenta where synchrotron losses
become important, the electron distribution function has at any given time
the form
\begin{equation} f_\mathrm{e}(p)= K_\mathrm{ep} f(p)  
\label{eq2}
\end{equation} 
with a factor $K_\mathrm{ep}$ that is of the order of $10^{-2}$ for the
average CRs in the Galaxy.

Clearly, from the point of view of injection/acceleration theory, we must
treat $K_\mathrm{ep}$ together with $B$ as free parameters, and $\eta$ as
well as $f_\mathrm{re}$ as theoretically not yet accurately calculable
parameters, to be {\it quantitatively} determined by comparison with
observations.

The electron distribution function $f_\mathrm{e}(p)$ deviates only at
sufficiently large momenta from Eq. (2) due to synchrotron losses, which
are taken into account by supplementing the ordinary diffusive transport
equation by a radiative loss term. The solution of the dynamic equations
at each instant of time yields the CR spectrum and the spatial
distributions of CRs and thermal gas. This allows the calculation of the
expected flux $F_{\gamma}^{\pi}(\epsilon_{ \gamma})$ of $\gamma$-rays from
$\pi^0$-decay due to $p-p$ collisions of CRs with the gas nuclei. The
choice of $K_\mathrm{ep}$ allows one then to determine the electron distribution
function in the energy region where losses can be neglected, and to
calculate the associated emission. Details about this calculation
are given in Berezhko et al. (\cite{bkv02}).

\section{Results and Discussion}

Tycho was a type Ia SN. Therefore we use typical SN Ia parameters in our
calculations:  ejected mass $M_\mathrm{ej}=1.4 M_{\odot}$, $k=7$, and a
uniform ambient ISM with hydrogen number density
$N_\mathrm{H}=0.5$~cm$^{-3}$ and temperature $T_0=10^4$~K. Following
Chevalier et al. (\cite{chevkr}) (see also Heavens
\cite{hea84}; Reynolds \& Ellison \cite{re92}; Aharonian et al.
\cite{aha01})  we in addition adopt a distance $d=2.3$~kpc, and a present
radius of $ \sim 4' $. These parameters are similar to those inferred by
Dickel \& Jones (\cite{dickel}) and Smith et al. (\cite{sm88}). Note that
SNR and CR dynamics are not sensitive to the precise value used for $T_0$,
because the shock structure is mainly determined by the Alfv\'{e}nic Mach
number.

We use an upstream magnetic field value $B_0=40$~$\mu$G, which is required
to provide the required shape of the synchrotron spectrum in the radio and
X-ray bands (see below).

The gas dynamic problem is characterized by the following length, 
time, and velocity scales: 
\[R_0=(3 M_\mathrm{ej}/ 4\pi \rho_0)^{1/3},~ t_0=R_0/V_0,~
V_0=\sqrt{2E_\mathrm{sn}/M_\mathrm{ej}},
\]
which are the sweep-up radius, sweep-up time and mean ejecta speed
respectively. Here $\rho_0=1.4 m_\mathrm{p}N_\mathrm{H}$ is the ISM mass
density, $m_\mathrm{p}$ is the proton mass.

According to Chevalier (\cite{chev82}), an analytical approximation to the
shock expansion law during the free expansion phase ($t<t_0$) is then
\begin{equation}
R_\mathrm{s} \propto E_\mathrm{sn}^{(k-3)/2k} \rho_0^{-1/k}
t^{(k-3)/(k-2)}, 
\label{eq9}
\end{equation}
which for $k=7$ gives
\begin{equation}
R_\mathrm{s}\propto (E_\mathrm{sn}^2/ \rho_0)^{1/7}t^{4/5}. \label{eq10}
\end{equation}
In the adiabatic phase $(t\,\,\raisebox{0.2em}{$>$}\!\!\!\!\!
\raisebox{-0.25em}{$\sim$}\,\, t_0)$ we have 
\begin{equation}
R_\mathrm{s}\propto (E_\mathrm{sn}/
\rho_0)^{1/5}t^{2/5}. \label{eq11}
\end{equation}
The observed mean expansion law of Tycho's SNR (Tan \& Gull \cite{tg85})  
is $R_\mathrm{s}\propto t^{\mu}$ with $\mu= 0.46 \pm 0.02$. This is
consistent with the average over the outer rim of the remnant, as obtained
by Reynoso et al. (\cite{rmg97}), even though the latter authors found
significant azimuthal variations of the expansion rate, indicating a
substantial nonuniformity of the SNR environment, as mentioned above. On
the whole therefore Tycho's SNR should be near the adiabatic phase (Strom
et al. \cite{strom}; Tan \& Gull \cite{tg85}; Reynoso et al.
\cite{rmg97}).

\begin{figure}
\centering
\includegraphics[width=7.5cm]{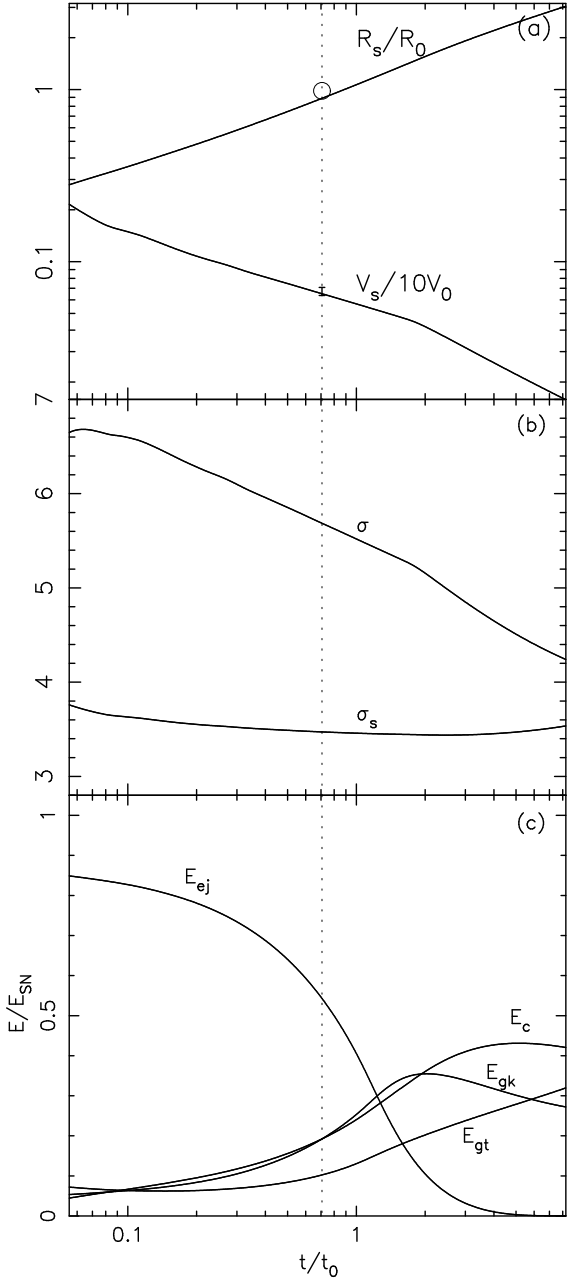}
\caption{(a) Shock radius $R_\mathrm{s}$ and shock speed
  $V_\mathrm{s}$; 
(b) total shock 
($\sigma$)
and subshock ($\sigma_\mathrm{s}$) compression ratios; (c) ejecta 
($E_\mathrm{ej}$), 
CR
($E_\mathrm{c}$), gas thermal ($E_\mathrm{gt}$), and gas kinetic 
($E_\mathrm{gk}$) energies
as a function of time.
Scale values are $R_0=2.72$~pc, $V_0=4402$~km/s,
$t_0=605$~years. The dotted vertical line marks the current 
epoch. 
The observed size and speed of the shock (Tan \& Gull \cite{tg85})
are shown as well.} \label{f1}
\end{figure}

The calculations together with the azimuthally averaged experimental data
are shown in Fig.\ref{f1}. An explosion energy $E_{\mathrm{sn}}=0.27\times
10^{51}$~erg is taken (in addition to the above ISM density, and the
ejecta mass) to fit the observed SNR size $R_\mathrm{s}$ and its expansion
rate $V_\mathrm{s}$.  This value is again in basic agreement with the
earlier determinations referred to above.

According to Fig.\ref{f1}a Tycho is indeed nearing the adiabatic phase.  To fit
the observed radio spectral index (see below) we assume a proton injection
rate $\eta=3\times 10^{-4}$. This leads to a significant nonlinear
modification of the shock which at the current age of $t=428$~yrs has a
total compression ratio $\sigma=5.7$ and a subshock compression ratio
$\sigma_\mathrm{s}=3.5$ (Fig.\ref{f1}b).

We should at this point ask the question, how sensitive these results are
to the specific choice of the overall parameters. These parameters are
partly the result of analyses of the thermal X-ray emission. The
corresponding X-ray models need to take into account of non-equilibrium
electron temperatures and non-equilibrium ionisation states behind the
outer shock as well as nonsolar chemical composition in the ejecta which
are in part heated by the reverse shock. As a consequence, different
workers do not obtain identical results. The differences to our set of
parameters is most pronounced in the results of Hamilton et al.  
(\cite{hss}) who derived the different values $E_{\mathrm{sn}}=0.7\times
10^{51}$~erg, $N_{\mathrm{H}}=0.28$~cm$^{-3}$, and $d=3.1$~kpc. One way of
comparing the consequences of such different inputs into our theory is to
check on their internal consistency. Taking into account that the linear
size $R_\mathrm{s} \propto d$, then it follows from Eq. (4) that
$E_{\mathrm{sn}} \propto N_{\mathrm{H}}^{1/2} d^{7/2}$ for a given source.  
Changing our parameters $N_{\mathrm{H}}$ and $d$ to
$N_{\mathrm{H}}=0.28$~cm$^{-3}$ and $d=3.1$~kpc, respectively, we obtain
$E_{\mathrm{sn}}=0.57\times 10^{51}$~erg for the Hamilton et al.  case,
not too different from their actual value for $E_{\mathrm{sn}}$. We can
also discuss the change expected for the resulting $\pi^0$-decay
$\gamma$-ray integral flux $F_{\gamma}^{\pi}$. Since in the strongly
nonlinear case, which is characterized by efficient CR acceleration, the
number density of accelerated CRs at relativistic energies is proportional
to $\rho_0V_\mathrm{s}^2$ (e.g. Berezhko et al. \cite{byk96}; Berezhko \&
V\"olk \cite{bv97}), the occupied volume goes like $R_\mathrm{s}^3$, and
the $\pi^0$-decay $\gamma$-ray production is proportional to the gas
density $\rho_0$, we have $F_{\gamma}^{\pi} \propto N_{\mathrm{H}}^2
R_\mathrm{s}^3 V_\mathrm{s}^2 d^{-2}$.  Inserting Hamilton et al.'s values
for $d$ and $N_{\mathrm{H}}$, the $\pi^0$-decay $\gamma$-ray flux would be
decreased to 0.77 times our value given in Fig.\ref{f6} below. Given our
value of the magnetic field, we can do the same for the electron-proton
ratio $K_\mathrm{ep}$, given the measured radio energy flux $\nu S_{\nu}$
(see Fig.\ref{f3}). The value of $K_\mathrm{ep}$ would be lowered by a
factor of 1.3 times our value. The IC $\gamma$-ray flux would not change
at all for given radio flux. The magnetic field value which we deduce has
a different character. We therefore defer its discussion to a later point
of this section.

We now continue with the nonthermal aspects of Tycho's SNR.

Given the proton injection rate the acceleration process is characterized
by a high efficiency in spherical symmetry:  at the current time
$t/t_0=0.7$ about 20\% of the explosion energy have been already
transferred to CRs, and the CR energy content $E_\mathrm{c}$ continues to 
increase
to a maximum of about 43\% in the later Sedov phase (Fig.1c), when
particles start to leave the source. As usually predicted by the
spherically symmetric model, such a CR acceleration efficiency is
significantly higher than required for the average replenishment of the
Galactic CRs by SNRs, corresponding to $E_\mathrm{c}\approx
0.1E_\mathrm{sn}$. As discussed above (V\"olk et al. \cite{voelk},
Berezhko et al. \cite{bkv02}), this requires a physical renormalization of
the number of hadronic CRs. We choose $f_\mathrm{re}=1/5$ here, in
agreement with the theoretical estimate above, and consistent with the
average Galactic CR replenishment.

With  this renormalization the CRs inside Tycho's SNR contain at the 
current epoch the total energy
\begin{equation}
E_\mathrm{c}\approx0.2f_\mathrm{re}E_\mathrm{sn}\approx10^{49}~\mbox{erg}.
\end{equation}
%
\begin{figure} \centering \includegraphics[width=7.5cm]{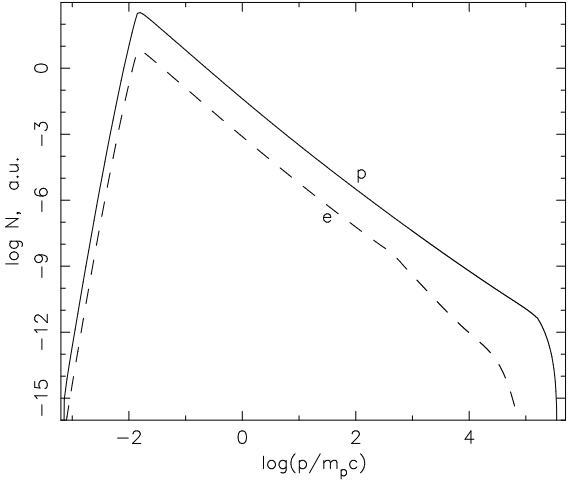}
\caption{The overall CR spectrum as function of momentum. 
Solid and dashed lines correspond to protons and electrons, respectively.} 
\label{f2} 
\end{figure}
The volume-integrated (or overall) CR spectrum 
 \begin{equation}
N(p,t)=16\pi^2p^2 \int_0^{\infty}dr r^2 f(r,p,t)  
\label{eq13}
\end{equation} 
is given in Fig.\ref{f2} (Note that only the nonthermal part of the spectrum is
pictured. That is why both the proton and electron spectrum have a maximum
above suprathermal proton energies. As discussed above, the spectrum
$N(p)$ is of course smoothly connected with the thermal distribution, not
given here.) Far above suprathermal energies it has, for the case of
protons, almost a pure power-law form $N\propto p^{-\gamma}$ over a wide
momentum range from $10^{-2}m_\mathrm{p}c$ up to the cutoff momentum
$p_\mathrm{max}=\epsilon_\mathrm{max}/c$, where
$\epsilon_\mathrm{max}\approx 2\times 10^{14}$~eV is the maximum CR energy
(Fig.\ref{f2}). This value $p_\mathrm{max}$ is limited mainly by
geometrical factors, which are the finite size and speed of the shock, its
deceleration and the adiabatic cooling effect in the downstream region
(Berezhko \cite{ber96}). Due to the shock modification the power-law index
slowly varies from $\gamma=2.2$ at $p\,\,\raisebox{0.2em}{$<$}\!\!\!\!\!
\raisebox{-0.25em}{$\sim$}\,\, m_\mathrm{p}c$ to $\gamma=1.8$ at
$p\,\,\raisebox{0.2em}{$>$}\!\!\!\!\!
\raisebox{-0.25em}{$\sim$}\,\,10^3m_\mathrm{p}c$.

The shape of the overall electron spectrum $N_\mathrm{e}(p)$ deviates from
that of the proton spectrum $N(p)$ at high momenta $p>p_\mathrm{l}\approx
10^3m_\mathrm{p}c$, due to the synchrotron losses in the downstream region
with magnetic field $B_\mathrm{d}\approx 240~\mu$G. This field is assumed
to be roughly uniform and equal to the potshock field $B_2$: $B_\mathrm{d}
\simeq \mathrm{B}_2 = \sigma_\mathrm{B} B_0$, where $\sigma_\mathrm{B}^2=
1/3 + (2/3) \sigma^2$ for a strongly turbulent field. We shall
approximately take $\sigma_\mathrm{B} = \sigma$ in the sequel.

The synchrotron losses become important for electron
momenta greater than
\begin{equation} 
\frac{p_\mathrm{l}}{m_\mathrm{p}c} \approx 
1.3 \left(\frac{10^8~\mbox{yr}}{t}\right)
\left(\frac{10~\mu\mbox{G}}{B_\mathrm{d}}\right)^2. \label{eq14}
\end{equation}
Substituting the SN age $t=428$~yr into this expression, we have
$p_\mathrm{l}\approx 800m_\mathrm{p}c$, in good agreement with the
numerical results (Fig.\ref{f2}).

The shock constantly produces the electron spectrum $f_\mathrm{e}\propto
p^{-q}$, with $q \approx 4$, up to the maximum momentum
$p_\mathrm{max}^\mathrm{e}$ which is much larger than $p_\mathrm{l}$.
Therefore, within the momentum range $p_\mathrm{l}$ to
$p_\mathrm{max}^\mathrm{e}$, the electron spectrum is $f_\mathrm{e}\propto
p^{-5}$ due to synchrotron losses, and the corresponding overall electron
spectrum is $N_\mathrm{e}\propto p^{-3}$.

The maximum electron momentum can be estimated by equating the 
synchrotron loss time and the acceleration time, that gives
\[
\frac{p_\mathrm{max}^\mathrm{e}}{m_\mathrm{p}c}
= 6.7\times 10^4 \left(\frac{V_\mathrm{s}}{10^3~\mbox{km/s}}\right)
\]
\begin{equation}
\hspace{1cm}\times
\sqrt{\frac{(\sigma-1)}{\sigma (1+\sigma_\mathrm{B} \sigma)}
  \left(\frac{10~\mu\mbox{G}}{B_0}\right)}. \label{eq17}
\end{equation}
At the current epoch $V_\mathrm{s}\approx 3100$~km/s which leads to a
maximum electron momentum $p_\mathrm{max}^\mathrm{e}\approx 6\times
10^4m_\mathrm{p}c$ in agreement with the numerical results (Fig.\ref{f2}).
\begin{figure}
\centering
\includegraphics[width=7.5cm]{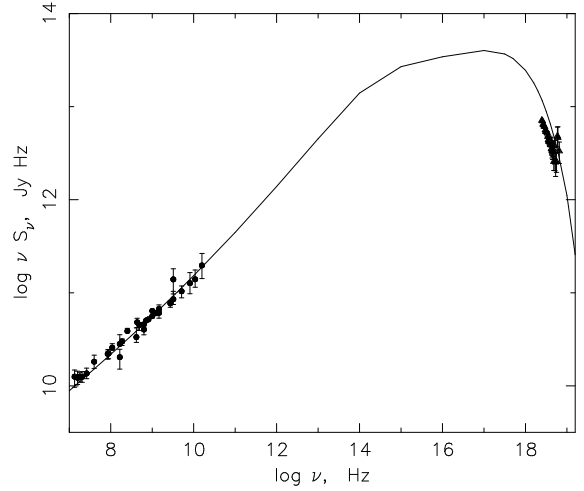}
\caption{Spectral energy distribution of the electron
synchrotron emission. The observed X-ray (Allen et al. \cite{agp99})  and
radio emission (Reynolds \& Ellison \cite{re92}) are shown.}
\label{f3}
\end{figure}
Overall, the parameters $K_\mathrm{ep}=4\times 10^{-3}$ and
$B_{\perp}=0.5B_\mathrm{d}= 120$~$\mu$G provide good agreement between the
calculated and the measured synchrotron emission in the radio to X-ray
ranges (Fig.\ref{f3}). This concerns first the deviation of the
radio-spectrum from the test particle form $S_{\nu} \propto \nu^{- 0.5}$.
At the same time, the steepening of electron spectrum at very high
energies due to synchrotron losses naturally yields a fit to the X-ray
data with their soft spectrum. Such a smooth spectral behavior is achieved
in a $40$~$\mu$G upstream field (see also Berezhko et al. \cite{bkv02}).
This is in clear contrast to the type of phenomenological fit by e.g.  
Reynolds (\cite{re98}) for SN 1006.

Since the total number of accelerated protons has to be renormalized by
the factor $f_\mathrm{re}=0.2$, the above number of accelerated electrons
then corresponds to the {\it renormalized} parameter
$K_\mathrm{ep}=2\times 10^{-2}$. This turns out to correspond reasonably
well to the canonical value of $10^{-2}$ of the electron to proton ratio
in the Galactic CRs. The total energy of accelerated electrons in Tycho's
SNR is approximately $E_\mathrm{c}^\mathrm{e}=2\times 10^{47}$~erg, as
estimated from the fit to the observed synchrotron emission (Fig.\ref{f3})
and our estimated value of $B_{\perp}$.
\begin{figure}
\centering
\includegraphics[width=7.5cm]{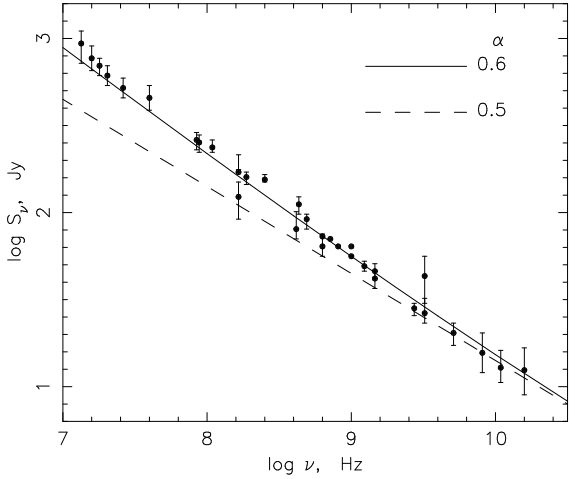}
\caption{Observed total radio flux of Tycho's SNR (Reynolds \& Ellison
\cite{re92}) as a function of frequency, with model spectra superimposed.
The solid curve represents our selfconsistent calculation, whereas the
dashed curve illustrates the shape of the spectrum that would be expected
in the test particle limit.}
\label{f4}
\end{figure}

In detail the radio data can be fitted with a power law spectrum
$S_{\nu}\propto \nu^{-\alpha}$, whose index $\alpha =0.607\pm 0.007$
(Reynolds \& Ellison \cite{re92}) is considerably larger than 0.5, as it
would correspond to an electron spectrum $N_\mathrm{e}\propto p^{-2}$
produced in the test particle limit by an unmodified shock with
compression ratio $\sigma=4$. For our choice of the proton injection rate,
$\eta=3\times 10^{-4}$, the shock is significantly modified by the
backreaction of the accelerated protons (see Fig.\ref{f1}b) with a total
present compression ratio $\sigma=5.7$ at the present epoch. At the same
time low energy electrons, with momenta
$p\,\,\raisebox{0.2em}{$<$}\!\!\!\!\! \raisebox{-0.25em}{$\sim$}\,\,
10m_\mathrm{p}c$ ($\epsilon_\mathrm{e}\,\,\raisebox{0.2em}{$<$}\!\!\!\!\!
\raisebox{-0.25em}{$\sim$}\,\, 10$~GeV) which produce synchrotron emission
at $\nu\,\,\raisebox{0.2em}{$<$}\!\!\!\!\! \raisebox{-0.25em}{$\sim$}\,\,
10$~GHz, are primarily accelerated at the subshock which only has a
compression ratio $\sigma_\mathrm{s}=3.5$. Therefore these electrons have
a steeper spectrum $N_\mathrm{e}\propto p^{-2.2}$ that leads to a radio
spectrum $S_{\nu}\propto \nu^{-0.6}$ that fits the experimental data quite
well (see Fig.\ref{f4}, where the synchrotron flux at radio frequencies is
presented).  The fact that the observed value of the radio power law index
$\alpha$ exceeds the value 0.5 is considered as an indication that the
shock is significantly modified.  Note that $\alpha=0.6$ is the average
value of the power-law index within the frequency range shown in
Fig.\ref{f4}. In fact due to the concave shape of the electron spectrum
the index slightly decreases with increasing frequency, with $\alpha
=0.61$ and $\alpha=0.56$ at the lowest and largest frequency respectively.  
We emphasize that in our view not only a strongly modified shock, which
produces a steep electron spectrum at energies $\epsilon_e <1$~GeV, but
also a relatively high upstream magnetic field strength $B_0=40$~$\mu$G,
compared with typical ISM values $B_0=5$~$\mu$G, and a corresponding
downstream value $B_{\perp}=120$~$\mu$G are unavoidably required to have
the energies of the radio emitting electrons in the steep part of their
spectrum $\epsilon_\mathrm{e}\,\,\raisebox{0.2em}{$<$}\!\!\!\!\!
\raisebox{-0.25em}{$\sim$}\,\, 1$~GeV. According to the model calculations
of Lucek \& Bell (\cite{lucb00}), the existing ISM field can be
significantly amplified near the shock by CR streaming to result in such
values. An upper bound for the field strength is discussed below.

\begin{figure}
\centering
\includegraphics[width=7.5cm]{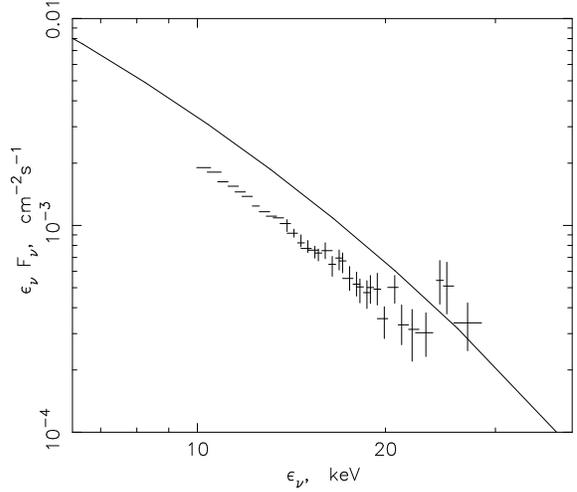}
\caption{Observed total integral hard X-ray flux of Tycho's SNR as a 
function of
  photon energy (Allen et al. \cite{agp99}), together with the theoretical 
synchrotron spectrum.}
\label{f5}
\end{figure}

We note that the necessity of strong nonlinear shock modification and
large values of the magnetic field strength in young SNRs to reproduce
their steep radio spectra was argued for the first time by Reynolds \&
Ellison (\cite{re92}).  The modified shock parameters and required
magnetic field $B_\mathrm{d}=10^{-4} - 10^{-3}$~G which they derived for
Tycho's SNR are consistent with our results.

The large perpendicular downstream magnetic field $B_{\perp}=120$~$\mu$G,
which leads to a substantial steepening of the electron spectrum at high
energies, also naturally provides a smooth cutoff in the synchrotron
spectrum $S_{\nu}(\nu)$ at frequencies $\nu> 10^{18}$~Hz, that correspond
to X-ray photon energies $\epsilon_{\nu}>5$~keV. As it can be seen from
Fig.\ref{f5} the spectral shape of the calculated integral flux
$\epsilon_{\nu} F_{\nu}(\epsilon_{\nu})$ is closely similar to what is
observed (Allen et al. \cite{agp99}); the X-ray data are just approaching
the cutoff region.  The small difference in the amplitude is not relevant,
because it could be easily reduced by a fine tuning of parameter values
which we do not attempt to achieve here.

The calculated IC emission off the CMB nonthermal bremsstrahlung (NB) and
the derived $\pi^0$-decay $\gamma$-ray fluxes are presented in
Fig.\ref{f6}, together with existing experimental data. Note that the
number of accelerated protons, which produce the $\pi^0$-decay
$\gamma$-rays, is renormalized by the factor $f_\mathrm{re}=0.2$ relative
to the calculation in spherical symmetry.

With this renormalization, the hadronic gamma-ray flux is just below the
HEGRA upper limit in Fig. \ref{f6}. The electron synchrotron -- and thus
also IC and NB -- fluxes have to remain the same, fixed by the radio and
X-ray observations.
\begin{figure}
\centering
\includegraphics[width=7.5cm]{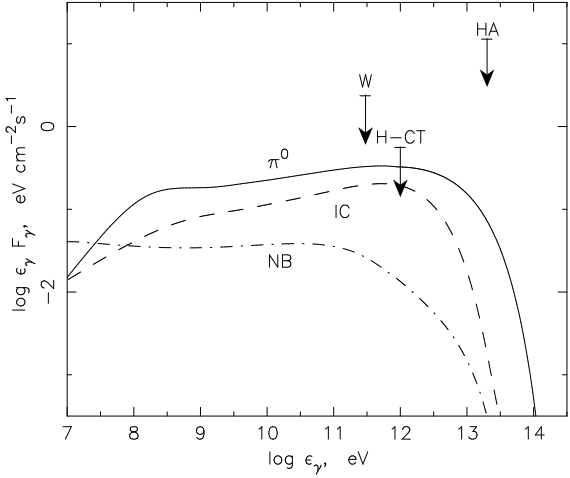}
\caption{IC (dashed line), NB(dash-dotted line), and $\pi^0$-decay (solid line)
$\gamma$-ray spectral energy distributions, as a function of $\gamma$-ray 
energy. The observed
$3 \sigma$ $\gamma$-ray upper limits (W -- Whipple (Buckley et al.,
\cite{buc98}), H-CT -- HEGRA IACT-system (Aharonian et al. \cite{aha01})),
and the 95\% confidence HA -- HEGRA AIROBICC upper limit (Prahl et al.
\cite{pp97}), are shown as well.}
\label{f6}
\end{figure}

According to the calculation, the hadronic $\gamma$-ray production exceeds
the electron contribution by a factor of about 2 at energies
$\epsilon_{\gamma}\,\,\raisebox{0.2em}{$<$}\!\!\!\!\!
\raisebox{-0.25em}{$\sim$}\,\, 1$~TeV, and dominates at
$\epsilon_{\gamma}>10$~TeV (Fig.\ref{f6}). The NB
is always small above the GeV range. We cannot say much about lower
energy NB because the electron distribution is not well known at those
lower energies, as discussed earlier. In the region of the $\pi^0$-decay
bump around 67.5 MeV, the hadronic gamma-ray spectrum may be
distiguishable from the electron IC spectrum. However, the corresponding
low particle energies are so far away from the "knee" in the Galactic CR
spectrum around $10^{15}$~eV that such a bump would not be a proof for
Tycho's SNR being a typical source of the nuclear CRs. In addition, the
steep background of the diffuse Galactic $\gamma$-ray flux would be hard
to remove for an extended source, as most nearby SNRs would be.

The $\gamma$-ray spectra produced by the electronic and hadronic CR
components have rather similar shapes at higher energies 10~GeV~$\simle
\epsilon_{\gamma}\simle~$1~TeV due to the synchrotron losses of the
electrons.  Therefore, the only clear observational possibility to
discriminate between the leptonic and hadronic contributions is to measure
the $\gamma$-ray spectrum at energies significantly higher than 1~TeV,
where these two spectra are predicted to be essentially different. The
detection of a substantial flux at energies $\epsilon_{\gamma}\simge
10$~TeV would provide direct evidence for its hadronic origin.

\subsection{$B$-field amplification possibilities; upper bound}

Strong shocks like those of young SNRs have their diffusive properties
dominated by self-excited waves from the CR streaming instability (Bell
\cite{bell78}). In quasilinear approximation the mean square field fluctuation
reaches a level $<\delta B^2>/4\pi \simeq (V_{\mathrm{a}}/V_{\mathrm{s}})1/2
\rho V_{\mathrm{s}}^2$, where $1/2 \rho V_{\mathrm{s}}^2$ is the upstream flow
energy density in the shock frame, and $V_{\mathrm{s}}/V_{\mathrm{a}}
\gg 1$ is the Alfv\'enic Mach number of the shock, with $V_{\mathrm{a}}=B/(4\pi
\rho)^{1/2}$ (McKenzie \& V\"olk \cite{mcv82}, see also Drury \cite{drury83}).
Nevertheless $<\delta B^2>/B^2 \simeq V_{\mathrm{s}}/V_{\mathrm{a}}$ which
implies that the rms wave field dominates the regular field.

Linear Alfv\'en waves constitute the long-wavelength elastic response of a
magnetized plasma to a transverse perturbation. In reality it is not clear why
the wave field could not grow even more strongly in the nonlinear regime due to
a nonlinear evolution of the CR streaming instability. Such a "plastic"  
response (V\"olk \& McKenzie \cite{voelk81}) makes the distinction between a
mean field and its fluctuations rather meaningless, effectively increasing the
magnetic field in which particles undergo spatial diffusion (V\"olk
\cite{voelk84}). Bell \& Lucek (\cite{bell2001}) have constructed an
interesting phenomenological model for such a nonlinear development, arguing
for a strong reduction of the effective diffusion coefficient. For strong
fluctuations the diffusion coefficient should come close to the Bohm limit
$\kappa \sim 1/3 r_g c$, with $r_g = cp/eB$, where $B$ is the effective
magnetic field strength. The maximum momentum to be reached in diffusive shock
acceleration is then proportional to $B$ for any given time the process is
assumed to operate.

How large the energy density of the effective field $B$ can actually become is
a complex question. However, an upper bound is given by imposing no other
constraint than the requirement that the Alfv\'en Mach number always remains
large compared to unity, so that the shock continues to be strong and to
produce the nonlinear field amplification in the first place. Requiring then
$V_{\mathrm{s}}/V_{\mathrm{a}} \leq 5$ we obtain $B_{\mathrm{ub}}^{2}/8\pi =
(1/25) 1/2 \rho V_{\mathrm{s}}^2$. Applying this to the shock of a type Ia SN,
we take for the interstellar field $B^{\mathrm{ISM}} = 5\mu$~G, $V_{\mathrm{s}}
= 5\times 10^3$~km/s during the sweep-up phase, and $B^{\mathrm{ISM}}/(4\pi
\rho)^{1/2} \simeq 14$~km/s, using our value of the ISM density ambient to
Tycho's SNR. Thus $B_{\mathrm{ub}}/B^{\mathrm{ISM}} \simeq 71$ as upper bound
for $B$.

Our estimate for the strength of $B$ in this paper consists in comparing the
observed synchrotron spectrum with a theoretical spectrum for the shock
accelerated electrons. We conclude that the $B$-value inferred is still well
within this upper bound, especially considering that it is actually the mean
square field value that counts.

An interesting corollary concerns the upper bound $p_{\mathrm{ub}}$ for the
proton cutoff momentum which is achieved at the end of the sweep-up phase.
Given that typical estimates yield $p_{\mathrm{ub}} \simeq 10^{14}$~eV in SNRs
for $B=B^{\mathrm{ISM}}$, we obtain for Tycho's SNR $p_{\mathrm{ub}} \simeq
7.1\times 10^{15}$~eV which corresponds to the so-called knee in the Galactic
CR spectrum.

Even if we were to assume a much higher value $B^{\mathrm{ISM}}/(4\pi
\rho)^{1/2} = 50$~km/s for the ISM Alfv\'en velocity, these conclusions would
not change qualitatively. This shows that at least in type Ia SNe the knee
energy is the upper bound for diffusive shock acceleration.

\section{Summary}

The kinetic nonlinear model for CR acceleration in SNRs has been applied
in detail to Tycho's SNR, in order to compare theoretical results with the
recently found very low observational upper limit for the TeV $\gamma$-ray
flux. We have used stellar ejecta parameters $M_\mathrm{ej}=1.4M_{\odot}$,
$k=7$, distance $d=2.3$~kpc, and ISM number density
$N_\mathrm{H}=0.5$~cm$^{-3}$.  A total hydrodynamic explosion energy
$E_\mathrm{sn}=0.27\times 10^{51}$~erg was derived to fit the observed
size $R_\mathrm{s}$ and expansion speed $V_s$ which are determined by the
ratio $E_\mathrm{sn}^2/N_\mathrm{H}$.  Even though the distance to the
object is not very well known, the set of parameters has been shown to be
internally consistent, and the predictions for the radio the and
$\gamma$-ray fluxes are quite robust with respect to different parameter
values in the literature.

A rather high downstream magnetic field strength $B_\mathrm{d}\sim 240
$~$\mu$G and a proton injection rate $\eta=3\times 10^{-4}$ are needed to
reproduce the observed steep and concave radio spectrum and to ensure a
smooth cutoff of the synchrotron emission in the X-ray region. We cannot
exclude that the required magnetic field strength, that is significantly
higher than the rms ISM value $5~\mu$G, might have to be attributed in
part to its nonlinear amplification near the SN shock by CR acceleration
itself. The evidence for efficient nucleonic CR production that comes from
the radio and X-ray data and leads to a strong shock modification, is
even more definite for Tycho's SNR than in the case of SN~1006 (Berezhko
et al. \cite{bkv02}).

We find that, after adjustment of the predictions of the nonlinear
spherically-symmetric model by a physically necessary renormalization of
the number of accelerated CR nuclei to take account of the
quasi-perpendicular shock directions in a SNR, quite a reasonable
consistency with most of the observational data can be achieved. The
resulting nonthermal electron to proton ratio turns out to be consistent
with the observed ratio in interstellar space. The total $\gamma$-ray flux
at 1 TeV (with the $\pi^0$-decay component exceeding the IC component)
comes out to be slightly larger than the most restrictive observational
upper limit from the HEGRA experiment. Given the remaining uncertainties
in the basic astronomical parameters of Tycho, in particular those coming
from analysis of the thermal X-ray data, we do not consider this as a
problem for the theory. It rather leads us to the prediction that
detectors with several times higher sensitivity, like MAGIC or VERITAS in
the Northern Hemisphere, should indeed detect this source above 100 GeV 
in $\gamma$-rays.

The expected $\pi^0$-decay $\gamma$-ray flux $F_{\gamma}^{\pi}\propto
\epsilon_{\gamma}^{-1}$ extends up to $\simge 30$~TeV, whereas the IC
$\gamma$-ray flux has a cutoff above a few TeV. Therefore the detection of
$\gamma$-ray emission at $\sim10$~TeV would in addition imply clear
evidence for a hadronic origin.

\begin{acknowledgements}
This work has been supported in part by the Russian Foundation for Basic
Research (grants 00-02-17728, 99-02-16325). EGB and LTK acknowledge the
hospitality of the Max-Planck-Institut f\"ur Kernphysik, where part of
this work was carried out. GPR acknowledges the receipt of a von Humboldt
fellowship.
\end{acknowledgements}

\end{document}